\documentclass[10pt,aps,prb,twocolumn,superscriptaddress,floatfix,showpacs,longbibliography]{revtex4-1}
\usepackage[utf8]{inputenc}
\usepackage{graphicx}
\usepackage{tabularx}
\usepackage[usenames,dvipsnames,table]{xcolor}
\usepackage[version=3]{mhchem}
\usepackage{braket}
\usepackage{upgreek}
\usepackage{hyperref}
\usepackage{amsmath, amssymb}
\usepackage[normalem]{ulem}
\usepackage{soul}

\definecolor{mark}{rgb}{0.85, 0.9, 1}
\definecolor{rred}{HTML}{CB4154}

\hypersetup{hidelinks}
\sethlcolor{mark}
\newcolumntype{Y}{>{\centering\arraybackslash}X}

\begin{document}

\title{Certification of quantum states with hidden structure of their bitstrings}
\author{O. M. Sotnikov$^{1}$, I. A. Iakovlev$^{1}$, A. A. Iliasov$^2$, M. I. Katsnelson$^{2,1}$, A. A. Bagrov$^{2,3,1}$ and  V.~V.~Mazurenko}

\affiliation{ Theoretical Physics and Applied Mathematics Department, Ural Federal University, Mira Str. 19, 620002 Ekaterinburg, Russia \\
$^2$ Institute for Molecules and Materials, Radboud University, 6525 AJ Nijmegen, the Netherlands \\
$^3$Department of Physics and Astronomy, Uppsala University, SE-75120 Uppsala, Sweden
}

\date{\today}

\begin{abstract}
The rapid development of quantum computing technologies already made it possible to manipulate a collective state of several dozen of qubits. This success poses a strong demand on efficient and reliable methods for characterization and verification of large-scale many-body quantum states. Traditional methods, such as quantum tomography, which require storing and operating wave functions on classical computers, become problematic to use in the regime of large number of degrees of freedom. In this paper, we propose a numerically cheap procedure to describe and distinguish quantum states which is based on a limited number of simple projective measurements in at least two different bases and computing inter-scale dissimilarities of the resulting bit-string patterns via coarse-graining. The information one obtains through this procedure can be viewed as a ``hash function'' of quantum state -- a simple set of numbers which is specific for a concrete many-body wave function and can be used for certification. By studying a number of archetypal examples, we show that it is enough to characterize quantum states with different structure of entanglement, including the chaotic quantum states. The connection of the dissimilarity to standard measures of quantum correlations such as von Neumann entropy is discussed. We also demonstrate that our approach can be employed to detect phase transitions of different nature in many-body quantum magnetic systems.
    
\end{abstract}

\maketitle

\section{Introduction}
Theoretical description of objects invisible to human eye represents one of the challenging but, at the same time, most intriguing problems in physics through its history. For example, despite incessant improvement of optical instruments and the ability to look into more and more distant corners of the Universe, in many cases one can conclude on the existence of a planet only in an indirect way by analyzing its tiny influence on the orbits of neighboring visible planets \cite{Batygin} and stellar brightness \cite{Vanderburg, Gilbert}. In the opposite limit of the atomic scale, the situation is even more complicated. When the object of our principle interest is a many-body quantum state, -- wave function or density matrix, -- we should conclude on its existence and properties indirectly on the basis of measurements. Moreover, in contrast to observation of celestial objects whose collective motion could be completely described with laws of classical mechanics, a measurement in quantum world does not provide a complete information about a system due to the uncertainty principle \cite{Heisenberg}, and characterizing quantum matter from such limited probes represents a non-trivial methodological and technical problem.

The conventional technique to analyze quantum state of a multi-component physical system is quantum tomography, which is based on the idea of complete \cite{tomo1} or partial \cite{tomo2} reconstruction of the wave function or density matrix from a number of measurements. Complexity of the tomographic procedure is mainly related to the number of qubits involved and the complexity of the quantum state itself, about which one might or might not have some prior expectations. In many cases, it could be non-trivial to choose a set of observables which is tomographically complete (or sufficient for partial reconstruction) and, at the same time, experimentally accessible \cite{tomo2}. The main fundamental limitation of quantum tomography is that one needs to store and manipulate the to-be-reconstructed quantum state on a classical computer, which makes characterization of systems that comprise more than a few dozens of qubits unfeasible. Taking into account that quantum states of 53 qubits can already be generated on modern quantum devices \cite{google2}, and a significant increase of this number is expected in the coming years \cite{IBMroadmap}, seeking an approach that overcomes this limitation appears to be a problem of high importance.

\begin{figure*}[t]
	\includegraphics[width=2\columnwidth]{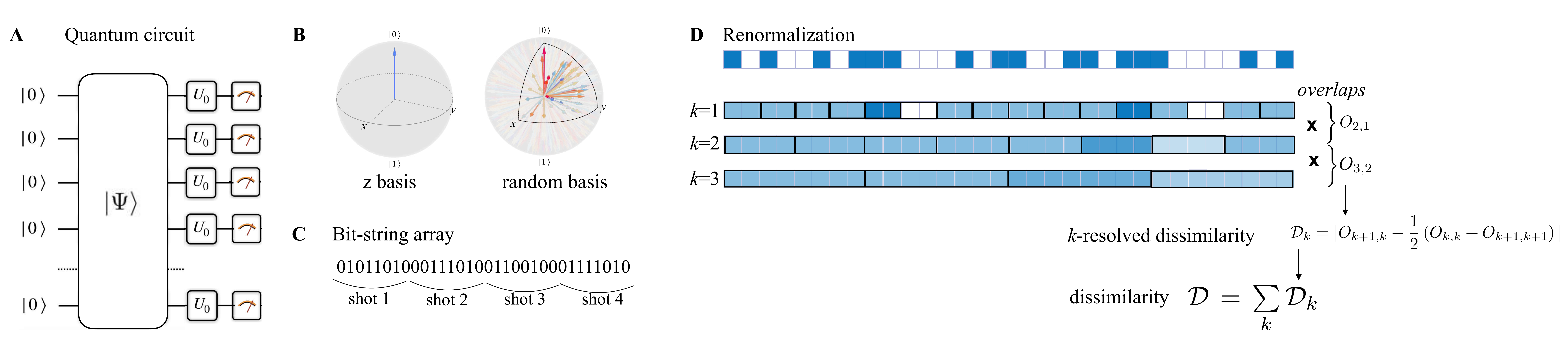}
	\caption{\label{fig:protocol} Protocol for computing dissimilarity of a quantum state. ({\bf A}) First, one prepares a state on a quantum device and chooses the measurement basis by applying rotational gates $U_{0}$ to individual qubits. ({\bf B}) In this paper, we work with $\sigma^z$ and random bases whose Bloch sphere representations are shown in the picture. We say that the set of measurements is performed in a random basis if, for each shot of measurement, a random vector belonging to the highlighted sector of the Bloch sphere is uniformly sampled and the corresponding parameters of gate $U_0$ are applied. ({\bf C}) A number of measurements is performed and their outcomes -- bitstrings of length $N$ -- are then stacked together in a one-dimensional binary array of length $N\times N_{shots}$ that serves as a classical representation of the quantum state. ({\bf D}) The array is coarse-grained in several steps (indexed with $k$). Different schemes can be employed, but here we use plain averaging with fixed filter size $\Lambda$. In the picture, blue and white squares in the top line correspond to ``0'' and ``1'' bits in the array shown in ({\bf C}), and black rectangles depict the blocks where averaging occurs at every step of coarse-graining. Overlap-based dissimilarities ${\cal D}_k$ between subsequent arrays are computed and summed up to the overall dissimilarity $\cal D$. See Methods section for more details.}
\end{figure*}

A natural way to reduce the memory required for state reconstruction is to store it in an implicit form of a compact variational ans\"atz. One of the most promising approaches of this kind is the recently proposed neural-network version of quantum tomography~\cite{tomography, Carrasquilla-generative}, which represents the wave function as a Neural Quantum State\cite{Carleo} and reconstructs it via the learning procedure. While this approach has many benefits such as very high expressibility of neural-network ans\"atze
\cite{XLi, Westerhout}, it does not resolve all the problems of quantum tomography. Some quantum states, such as defined by wave functions with random or uniform distributions of amplitudes over the Hilbert space basis, require exponentially large number of measurements (of the order of the Hilbert space dimension) for reconstruction. The situation cannot be improved by employing neural networks, since there are no features that the neural network can detect in the measured data, learn and generalize \cite{tomography}. Here, a natural question arises: can one somehow by-pass the resource-consuming routine of conventional quantum tomography at least in certain contexts?
A typical problem, when there is a chance not to get engaged in this procedure, is certification of a state prepared on a quantum information processing device. In this case, there are strong prior expectations of what this state should be. Thus, instead of its complete reconstruction, one could hope to read out simple signature serving as a fingerprint of the many-body state, - in a spirit similar to hash functions in computer science\cite{hash1,hash2}, - to make sure that the state is, with high probability, indeed the correct one (see Ref.\onlinecite{Wilczek-overlapping} for the usage of hash functions in quantum tomography).

In this paper, we introduce such a signature that can be constructed by means of a reasonable number of simple von Neumann measurements of the quantum state and does not require computing correlation functions. Ideologically, this can be viewed as going along the line of the very recent approach of classical shadow tomography \cite{Aaronson-shadow, Preskill-shadow}, though the signature we employ is different. To accomplish that, we heavily rely on the concept of multi-scale structural complexity of classical patterns that has been recently defined by some of the authors of this paper\cite{complexity}. To avoid possible terminological confusions with the well-established notion of quantum complexity, here we call it dissimilarity (since it is based on counting how much different spatial scales of an object differ from each other). The detailed description of the protocol is given in the Methods section, and here we outline the main idea.

Assume, we have access to a many-body quantum state. To do benchmark tests, in this paper we use both numerical wave functions (e.g., resulting from exact diagonalization) and physical quantum states generated on the IBM quantum simulator \cite{IBM}.
With no loss of generality, we will be considering spin-$1/2$ systems. A single-shot projective measurement of such a state results in a string of bits of length $N$, -- measured spin projections on a chosen direction: $|S_i\rangle = |0110\dots010\rangle$ ($0$ for spin-down and $1$ for spin-up), -- where $N$ is the number of qubits. Performing the measurement many times (denote this number with $N_{shots}$) and collecting the outcomes in a string, we obtain a bit-string array of length $L=N\times N_{shots}$.
This array can then be viewed as a one-dimensional pattern, and its inter-scale dissimilarity can be computed. For that, we do several steps of coarse-graining (we label the steps with index $k$) and for each pair of subsequent scales compute how distinct the corresponding coarse-grained strings are. The distinction is assumed to be large if overlap of arrays at two subsequent scales is small. For two neighboring scales, we call these measures partial dissimilarities, ${\cal D}_k$, and their sum over all scales ${\cal D}=\sum_k {\cal D}_k$ gives the total inter-scale dissimilarity. Different schemes of coarse-graining can be employed, and here we resort to the simplest option: we fix filter of width $\Lambda$ (usually $\Lambda=2$), and at step $k$ we substitute all the pixels within a window of size $\Lambda^k$ with the average value of pixels in this window at the previous step,  Fig.\ref{fig:protocol}.
Despite probabilistic nature of the measurements, in all the tested cases dissimilarity turns out to be a statistically robust signature of the state.

\begin{figure*}[!ht]
	\includegraphics[width=1.7\columnwidth]{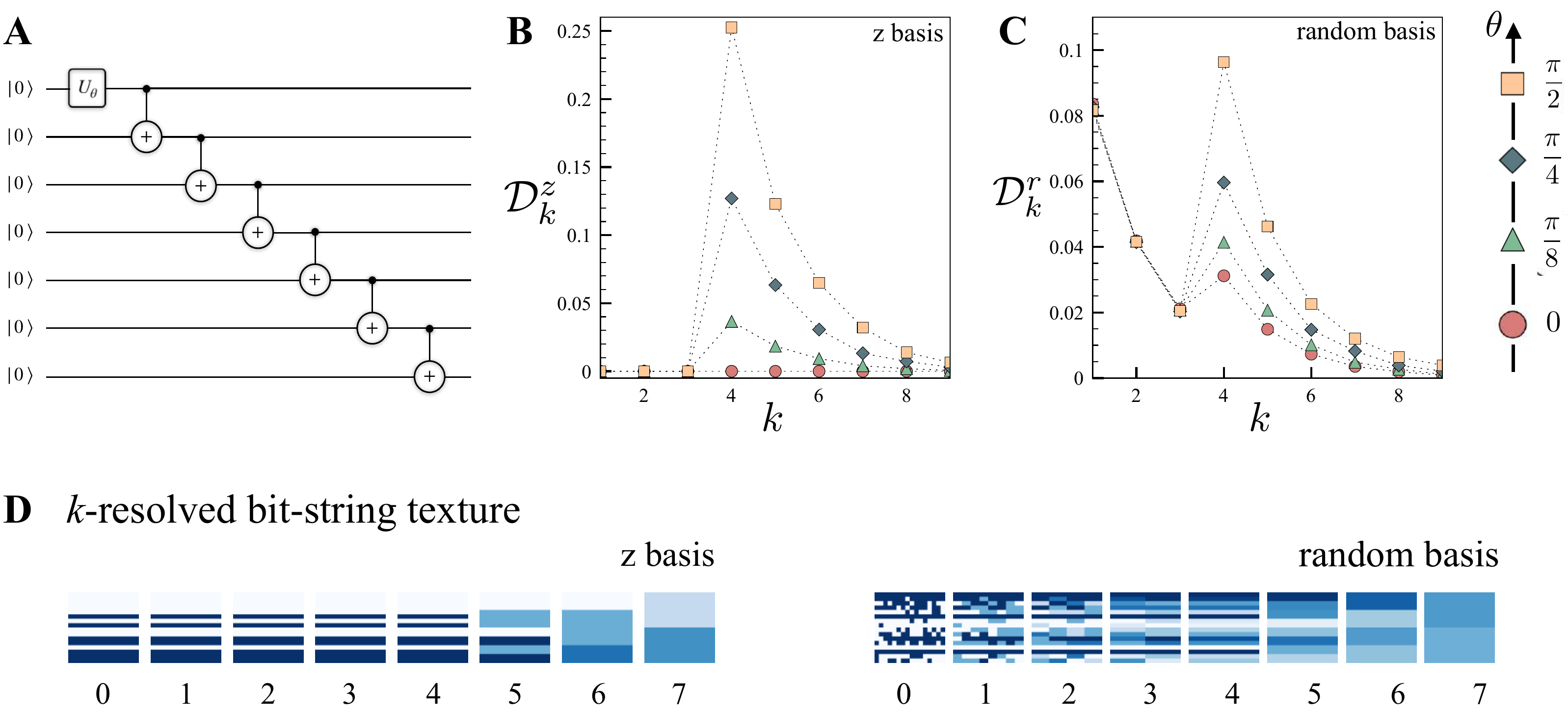}
	\caption{\label{fig:cat} ({\bf A}) Quantum circuit generating Schr\"odinger cat states. ({\bf B}, {\bf C}) Partial dissimilarities ${\cal D}_k$ of 16-qubit Schr\"odinger cat states calculated in the $\sigma^{z}$ and the random bases correspondingly. Here, $\Lambda = 2$. ({\bf D}) Visualization of bit-string arrays. In these images, individual bitstrings are horizontal lines of $16$ bits that are stacked vertically in an array ($16$ strings in total). Left picture shows an example of array sampled from a cat state with $\theta = \frac{\pi}{2}$ in the $\sigma^{z}$ basis, and the right one -- measured in the random basis. Here $k = 0$ represents texture of the measured array {\it per se}, and $k>0$ show its evolution upon coarse-graining.}
\end{figure*}

If this procedure is performed in a single basis, it does not reveal any information on the phase structure of the quantum state, since measurement outcomes are defined solely by probability distribution on the Hilbert space basis $|\Psi(S_i)|^2$. Also, unique characterization of a many-body quantum state with a single number is clearly impossible.
However, if such bit-string arrays are constructed in two or more different Hilbert space bases, one obtains a sequence of numbers that implicitly contains information on both amplitude and phase structure of the state. The more bases are involved, the less is it likely that two different quantum states would share the same dissimilarity signature (in a different context, the tomographic advantage of using several bases was discussed in Ref.\onlinecite{basis-dependent}).

In this paper, we do not go beyond measurements in two bases, and this seems enough to characterize several important families of quantum states. As a warm up, in Sec. \ref{sec:IIA} we consider the families of Dicke and Schr\"odinger cat states which have compact analytical representations, and demonstrate how the concept of bit-string inter-scale dissimilarity can be used for dimensional reduction and visualization of specific signatures of wave functions. We also reveal the connection between the dissimilarity measure and the von Neumann bipartite entanglement entropy which plays a central role in quantum information theory. In Sec. \ref{sec:IIB}, we test our approach by using it for certification of random quantum states characterized by complete delocalization in the Hilbert space, which we do both numerically and analytically. We also show that the proposed approach scales nicely and requires the same experimental efforts to certify 16-qubit and 53-qubit states. In Sec. \ref{sec:IIC}, using the transverse-field Ising model and the Shastry-Sutherland model as playgrounds, we show that the inter-scale dissimilarity can be used as a universal tool for detecting quantum phase transitions in many-body systems. Finally, in Sec. \ref{sec:IID} we discuss how the concept of inter-scale dissimilarity can be used for dimensional reduction and visualization of many-body quantum states.

\section{Results}
\subsection{Notable entangled quantum states}
\label{sec:IIA}

To demonstrate the idea of bit-string arrays and inter-scale dissimilarity, we begin with the Schr\"odinger cat states defined by superposition of merely two basis vectors in the Hilbert space
\begin{eqnarray}
|\Psi_{\theta} \rangle = {\rm cos}(\frac{\theta}{2}) |0\rangle^{\otimes N} + {\rm sin}(\frac{\theta}{2}) |1\rangle^{\otimes N}.
\end{eqnarray}
Parametrized by angle $\theta$, this family of states interpolates between trivial product state $|0\rangle^{\otimes N}$ at $\theta=0$ and the famous Greenberger-Horne-Zeilinger (GHZ) state $\Psi_{\rm GHZ} = \frac{1}{\sqrt{2}}(|0\rangle^{\otimes N} + |1\rangle^{\otimes N})$ at $\theta = \frac{\pi}{2}$.
These states can be realized with quantum circuit \cite{Cat_circuit} shown in Fig.\ref{fig:cat} A. First, with rotational gate $U_{\theta}$ one prepares ${\rm cos}(\frac{\theta}{2})|0\rangle + {\rm sin}(\frac{\theta}{2})|1\rangle$ state of one of the qubits in the system and takes it as a control qubit to perform controllable-NOT operation on the second qubit. This operation results in a two-qubit entangled state ${\rm cos}(\frac{\theta}{2})|00\rangle + {\rm sin}(\frac{\theta}{2})|11\rangle$. Repeating it $N-1$ times, one eventually entangles all the qubits and obtains the target Schr\"odinger cat state.

In $\sigma_z$-basis, projective measurements of such states can only result in either $0000\dots0$ or $1111\dots1$ bitstring. Clearly, first steps of coarse-graining affect only internal content of individual bit-strings of length $N$, where it simply maps $0000\dots0\rightarrow0000\dots0$ and $1111\dots1\rightarrow1111\dots1$. Thus the randomly assembled array of bitstrings remains intact, and partial dissimilarities ${\cal D}_k \equiv 0$ for $k$ such that $\Lambda^k < N$ (for $k<4$ when we take $N=16$ and $\Lambda=2$). At $\Lambda^k \geq N$, the coarse-graining flow starts mixing individual bitstrings, and non-trivial contributions to the dissimilarity emerge. 
In random basis, ${\cal D}_k$ take finite values at all scales $k$, though due to the trivial structure of basis vectors defining $\Psi_{\theta}$ partial dissimilarities do not depend on $\theta$ at $\Lambda^k < N$. 

Importantly, each state reveals a distinct set of ${\cal D}_k$ which can be used to distinguish states from each other. Schr\"odinger cat states are the simplest example of many-body entangled wave functions, but in what follows we will show that the same idea can be exploited when dealing with much more complex states.
It has to be stressed out one more time that, while individual bitstrings are assembled into array in a random order set by outcomes of consequent projective measurements, the partial dissimilarities and their total sum are robust upon repeatedly performing the set of measurements.

Another type of entangled states that are instructive to consider is the family of Dicke states \cite{Dicke},
\begin{eqnarray}
|\Psi_D \rangle = \frac{1}{\sqrt{C^N_D}} \sum_{j} P_{j}(|0\rangle^{\otimes N-D} \otimes |1\rangle^{\otimes D}),
\label{eq:Dicke_wf}
\end{eqnarray}
where the sum goes over all possible permutations of qubits. By increasing $D$ from 1 to $\frac{N}{2}$, one increases the number of basis vectors involved into the quantum state. Recently, these states have been experimentally realized \cite{Dicke_exp1,Dicke_exp2}, and their verification \cite{Dicke_exp3} is a challenging task if the number of qubits is large \cite{Dicke_verify}. As a proof of concept, in this paper we study Dicke states of 16 qubits, and initialize them on quantum simulator using the Least Significant Bit procedure \cite{LSB}.

\begin{figure}[b]
	\includegraphics[width=1\columnwidth]{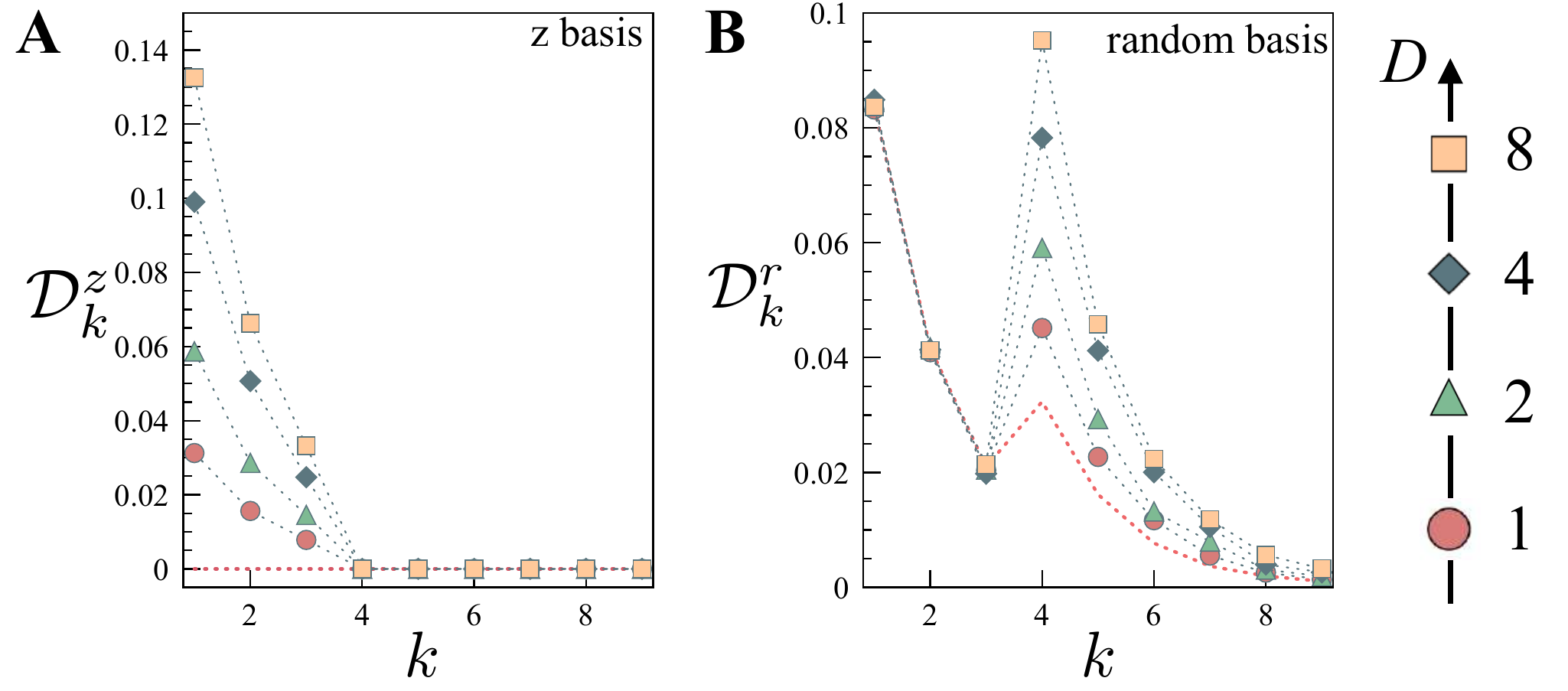}
	\caption{\label{fig:Dicke} Partial dissimilarities of Dicke states with different $D$ index calculated in the $\sigma^z$ ({\bf A}) and the random ({\bf B}) bases. The trivial state ($|0\rangle^{\otimes 16}$) profiles (dashed red lines) are given for comparison.}
\end{figure}

Partial dissimilarities of $16$-spin Dicke states computed in $\sigma_z$-basis and in the random basis with filter size $\Lambda=2$ are shown in Fig. \ref{fig:Dicke}. One can see that two different bases encode information about two ranges of scales. For any given parameter $D$, when bit-string arrays are constructed from measurement in the $\sigma^z$ basis, ${\cal D}_k$ take non-zero values only for $k < 4$, which follows from the fact that all the Hilbert space basis vectors possessing non-zero amplitudes have equal amount of spin-up entries, and after $4$ steps of averaging every bit string reduces to exactly the same number, and all the patterns are destroyed. Contrary, in the random basis, states with different $D$ can be distinguished from ${\cal D}_k$ at larger spatial scales, $k \geq 4$.

Since both families of states smoothly interpolate between regimes of low and high entanglement, it is interesting to study if there are any relations between the introduced measure of inter-scale dissimilarity and quantum correlations. To do that, we consider the von Neumann entanglement entropy
\begin{gather}
S(\rho_{\rm A}) = - {\rm Tr}_{\rm A} \rho_{\rm A} {\rm log}_2(\rho_{\rm A}), \\
\rho_{\rm A} = {\rm Tr}_{\rm B} \rho_{\rm AB}, \nonumber
\label{eq:vonentropy}
\end{gather}
where the system is divided into two equal parts $A$ and $B$ of $N/2$ qubits, compute its dependence on either $\theta$ or $D$ (depending on the family), and plot it alongside the inter-scale dissimilarity of bit-string arrays computed in the $\sigma_z$ basis.

\begin{figure}[!ht]
	\includegraphics[width=1\columnwidth]{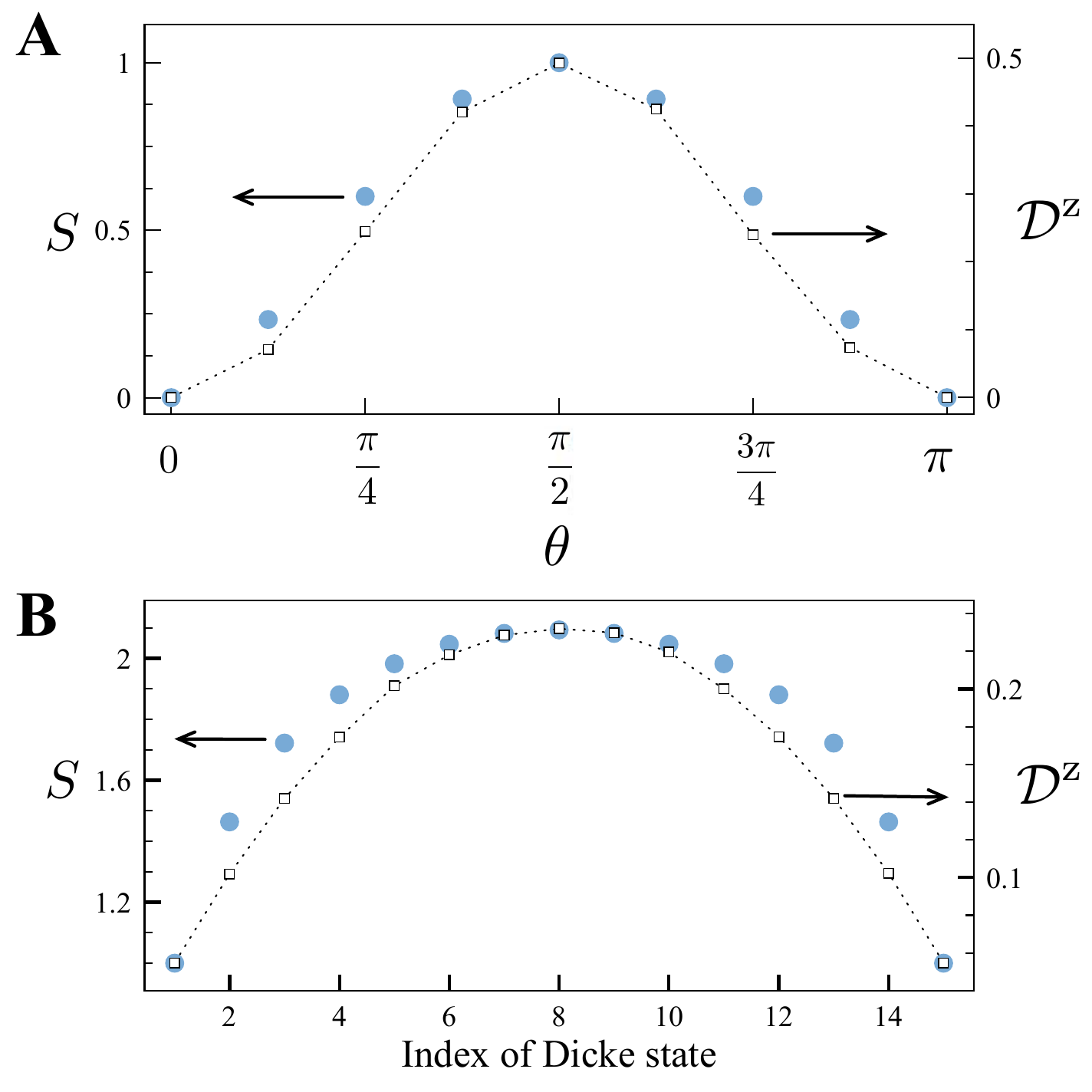}
	\caption{\label{fig:Cat_ent} Top: entanglement entropy $S$ (blue circles) and overall dissimilarity ${\cal D}^z$ (white squares) of the Schr\"odinger cat states as functions of angle $\theta$. Bottom: the same characteristics of the Dicke states as functions of index $D$.}
\end{figure}

\begin{figure*}[!ht]
	\includegraphics[width=2\columnwidth]{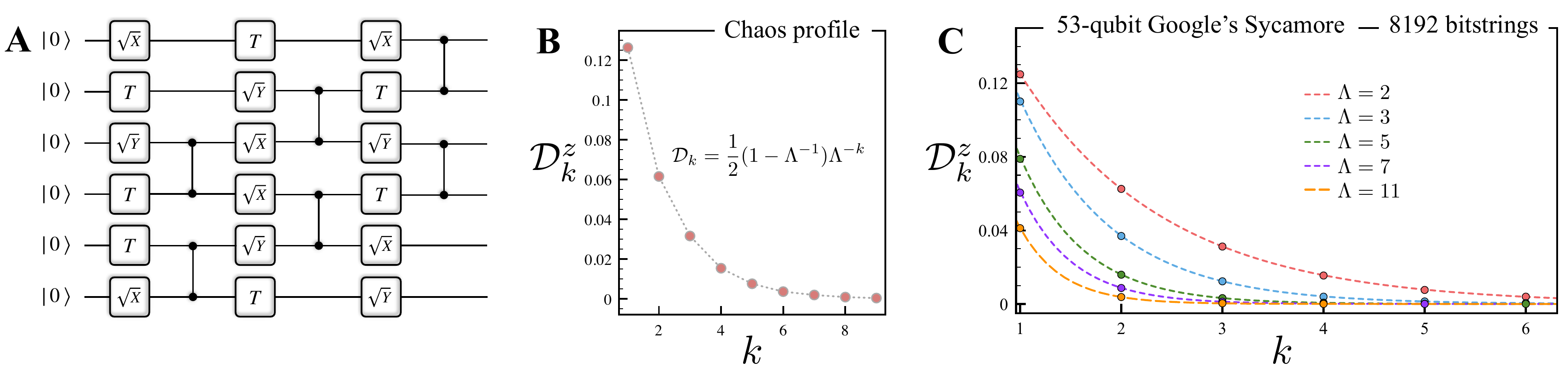}
	\caption{\label{fig:Chaos} ({\bf A}) Fragment of a quantum circuit generating chaotic quantum state according to the protocol proposed in Ref.\onlinecite{Richter}. ({\bf B}) Partial dissimilarities (red circles) of bit-string arrays resulting from 8192 projective measurements of a 19-layer-deep quantum chaotic circuit with 16 qubits in the $\sigma_z$ basis. Here, filter size $\Lambda=2$. Dashed line shows the analytical fit with \eqref{eq:Dk_fit}. ({\bf C}) Partial dissimilarities of bit-string arrays resulting from 8192 projective measurements of the state produced by the 53-qubit Sycamore quantum processor by Google. These data were taken from Ref.\onlinecite{google2}, and different filter sizes $\Lambda$ were used to compute ${\cal D}_k$. Dashed lines show the analytical fits.}
\end{figure*}

The result is shown in Fig. \ref{fig:Cat_ent}. While the Dicke and the Schr\"odinger cat states are quite different in the regard that variation of parameter $D$ modifies the structure of the wave function support in the Hilbert space basis, and $\theta$ only changes the balance between two basis vectors bearing non-zero amplitudes, in both cases dissimilarity nicely captures dependence of entropy on the parameters labeling the state within the family. Although the precise analytical correspondence between these two concepts is still to be revealed, it could be a good indication that it is possible to employ dissimilarity to estimate entanglement entropy, which is generally very difficult to reconstruct from experimental measurements, especially when dealing with multi-qubit systems inaccessible to quantum tomography. In a certain way, it is similar to the approach proposed in Ref.\onlinecite{Heyl-recognition}, where it was shown that, with the help of neural networks, entanglement can be reconstructed from visual pattern representations of quantum states.  

\subsection{Random quantum states}
\label{sec:IIB}
Our next goal is to demonstrate that the dissimilarity measures can serve as a signature not only of highly structured states with simple analytical representations, but of rather generic many-body states. To do that, we consider Haar-random wave functions uniformly sampled from the Hilbert space and characterized by the Porter-Thomas distribution of bit-string probabilities $p=|\langle x_1, \dots x_N|\psi \rangle|^2$ that have recently been used to demonstrate quantum supremacy\cite{google2}. These states play an important role in studying  quantum chaos theory \cite{chaos}, quantum information theory\cite{Page,qtheory} and information processing, including research domains of superdense coding of quantum states \cite{superdense} and data hiding \cite{qhide1,qhide2}, and even transport phenomena \cite{Richter}. While complete tomography of a given random state is an extremely complicated task since the minimal number of measurements to be performed to reconstruct a random quantum state should be of order of the Hilbert space dimension \cite{tomo1, tomography}, here we show that to certify if a state belongs to the Haar-random class one can resort to computing inter-scale dissimilarities of relatively short bit-string arrays.

As it was shown in Refs.\onlinecite{google1,google2,Richter,random}, random quantum states can be initialized with shallow pseudo-random circuits that can differ in the number and types of gates, and practical realization of these circuits on a real quantum device depends on its architecture. In this work, we generate random quantum states of a 16-qubit system on the IBM quantum simulator with the protocol proposed in Ref.\onlinecite{Richter}, which guarantees an accurate approximation of the Haar-random state with a compact circuit shown in Fig.\ref{fig:Chaos} A. More specifically, the circuit is formed in cycles, each having one- and two-qubit-gate layers. Within the first layer, for each qubit in system one randomly chooses from $\sqrt{X}$ , $\sqrt{Y}$ and $T$ gates, where $\sqrt{X}$ ($\sqrt{Y}$) are $\pi/2$ rotations around the $x$-axis ($y$-axis) of the Bloch sphere, and the non-Clifford gate $T = {\rm diag} (1,e^{i\pi/4})$. In turn, the second layer comprises controlled-Z gates, ${\rm diag}(1,1,1,-1)$, whose topology is randomly chosen from the set of configurations with fixed couplings between qubits, as described in Ref.\onlinecite{Richter}.   

In both $\sigma_z$ and random bases, the inter-scale partial dissimilarities of the array generated by sampling $8192$ bit-strings from a random quantum state follow the same decaying profile, Fig.\ref{fig:Chaos} B. Such a profile is a robust signature of typical Haar-random states. It remains the same even in the presence of noise and gate imperfections which we simulated by using the noise models provided by IBM with parameters corresponding to real quantum devices Paris and Montreal. It can be shown that, for a chosen filter size $\Lambda$, the dependence of ${\cal D}_k$ on the step index $k$ obeys a simple analytical law in the averaging coarse-graining scheme:
\begin{eqnarray}
{\cal D}_{k} = \frac{1}{2} (1 - \Lambda^{-1}) \Lambda^{-k}.
\label{eq:Dk_fit}
\end{eqnarray}
To derive this law from Eq.\ref{eq:Dissimilarity}, the central limit theorem must be employed as elaborated in the Methods section. This dependence is easy to reconstruct from a limited number of simple projective measurements, and it serves as a signature of the class of typical Haar-random states.

To go beyond the simple 16-qubit case and perform an ultimate test of the method, we have applied it to the real experimental data generated on the Google Sycamore quantum processor\cite{google2}. For systems of 16, 32 and 53 qubits, we have taken 8192 bitstrings measured in the $\sigma^z$ basis and calculated partial dissimilarities, which turned out to perfectly fit Eq.\eqref{eq:Dk_fit}. The result for the prominent example of 53-qubit system is presented in Fig.\ref{fig:Chaos} {\bf C}. 

In a real-world scenario, the bit-string arrays are clearly a subject to the gate errors and other sources of noise, and we have to understand how these imperfections are reflected in the dissimilarity signatures of the state. Previous studies \cite{google1,google2} have demonstrated that random quantum states are hypersensitive to the gate errors, which is considered to be a defining property of quantum chaos. When the error rates increase, the distribution of probabilities of the bitstrings generated by a random circuit deviates from the Porter-Thomas law $Pr(p)=2^{N} e^{-2^{N} p}$ and converges to equal probabilities of all the bitstings: $Pr(p)=\delta(1/2^{N} -p)$. To quantify this deviation, the authors of Refs.\onlinecite{google1,google2} have introduced the cross-entropy benchmarking procedure. It allows to estimate with a limited number of measurements how close a sampler -- a given quantum circuit -- to one of the two limiting cases: the ideal random quantum circuits with Porter-Thomas distribution of probabilities and uniform sampler with identical probabilities $p(x_1,\dots x_n)=2^{-N}$. In this respect, it is naturally to ask: can one distinguish between outputs of quantum circuits with the Porter-Thomas and the uniform probability distributions by calculating the inter-scale dissimilarity? 

To answer this question, we prepared a quantum circuit consisting of only the Hadamard gates that generates a 16-qubit state with uniform probabilities in the $\sigma^z$ basis: $|X\rangle  = \left({\rm H}|0\rangle\right)^{\otimes 16}$. Each qubit is then in the superposition $(|0\rangle + |1\rangle)/\sqrt{2}$. The obtained dissimilarity profile of the generated uniform state fully coincides with that obtained for random quantum circuits (Fig.\ref{fig:Chaos} B), with the overall dissimilarity ${\cal D}^z=0.25$. Thus, from $\sigma^z$ basis measurements we cannot distinguish these two states that are fully delocalized in the Hilbert space. However, in the random basis they have different profiles of ${\cal D}_k$ and overall $\cal D$. While the chaotic quantum circuit is characterized by an isotropic character of the dissimilarity that is independent on the measurement basis, the $|X\rangle$ state in the random basis reveals its trivial nature and the resulting dissimilarity ${\cal D}^r$ = 0.204 coincides with that obtained for $|0\rangle^{\otimes 16}$. This suggests that the inter-scale dissimilarity can be used to quantify deviations from a truly chaotic quantum states, which would be interesting to verify experimentally.

\subsection{Phase transitions in magnetic systems}
\label{sec:IIC}
Since the inter-scale bit-string dissimilarity appears to be a rather unique signature of many-body state, it is natural to expect that it should be sensitive to crossing phase boundaries in the parametric spaces of many-body quantum systems. If so, one can hope that it can be used as a sensitive indicator of phase transitions and directly used for constructing quantum phase diagrams, 
which is a crucial task in understanding phenomenology of correlated materials and designing new materials.  The common practice is to distinguish different phases of a quantum or classical many-body system by calculating the order parameter \cite{Imada, Wang} and low-order correlation functions such as susceptibility, scalar chirality and others. However, in many cases devising the order parameter is a non-trivial analytical problem, especially in the case of topological phases \cite{Berry,quantum_skyrmion}. 
Besides that, a quantum system may have a rich variety of different electronic and magnetic phases depending on internal (interactions) and external (temperature, pressure, magnetic field) parameters, and there could be no universal operator that can probe the whole phase diagram.

To overcome this problem, a lot of effort has been put into designing alternative approaches based on neural networks \cite{Huber, Kim, Biamonte1}, unsupervised machine learning techniques \cite{qt-SNE}, and quantum information theory concepts \cite{Bagrov, Rey}. These methods usually rely on manipulating eigenstates of the quantum system on a classical computer, which puts natural limitations on the size of systems that can be studied in this way. Also, it can be time- and resource-demanding to conduct, e.g., the learning procedure. 

\begin{figure}[h]
	\includegraphics[width=1\columnwidth]{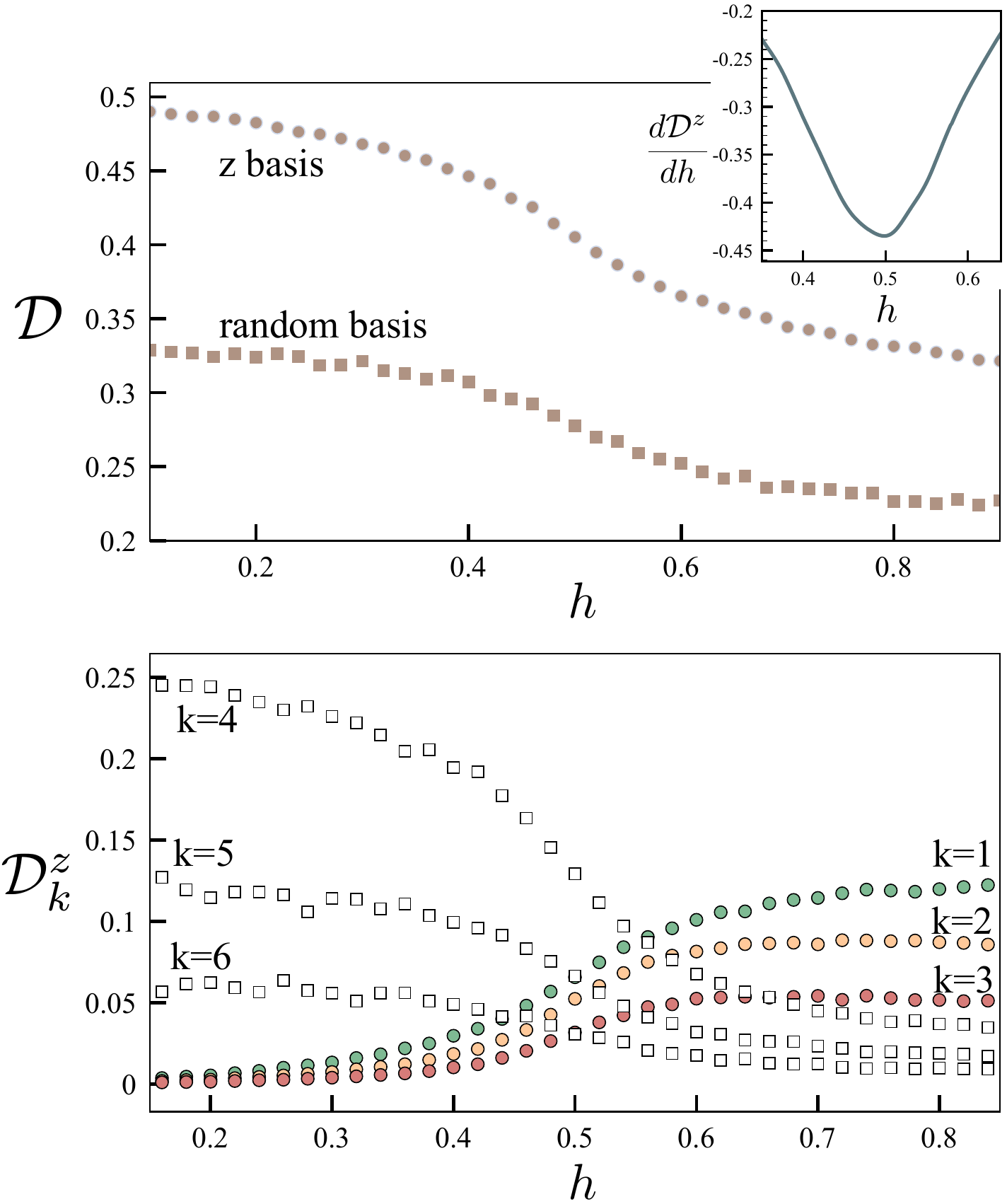}
	\caption{\label{fig:Ising} Top: dissimilarity of the Ising model ground state as a function of the transverse magnetic field in the $\sigma^z$ and the random bases; the inset shows derivative of the dissimilarity in the $\sigma^z$ basis with respect to $h$. Bottom: partial dissimilarities ${\cal D}_k$ in the $\sigma^z$ basis at different coarse-graining steps $k = 1\dots6$.}
\end{figure}

The progress in developing quantum simulators and quantum computing devices suggests a distinct way for large-scale representation of a quantum systems and analysis of their phase diagrams. Instead of solving the Hamiltonian numerically, one can imitate it in an, e.g., optical experiment. For example, by varying depth of the potential in optical lattices, one can change the ratio between hopping integrals and on-site Coulomb interaction in the simulated strongly-correlated electronic or bosonic system, and scan through its parametric space in this way. Recent advances in this field include simulation of the electronic metal-to Mott insulator transition \cite{AMazurenko} and destruction of the antiferromagnetic long-range order with temperature and doping \cite{AMazurenko1}. Analysis of such experiments is then conducted by means of a limited set of site-resolved measurements performed on the system, and the relevant information should be extracted from these measurements, whose number is much smaller than the Hilbert space dimension. We refer the reader to Refs.~\onlinecite{Demler, Khatami} for an interesting machine learning-based approach to the analysis of optical lattice experiments, and in what follows we discuss how the concept of bit-string arrays and their inter-scale dissimilarity can enter the game and aid reconstruction of phase diagrams of simulated quantum matter.

As some of us have shown in Ref.\onlinecite{complexity}, the classical prototype of inter-scale dissimilarity, - the structural complexity of patterns, - can be used to detect phase transitions in classical systems without any prior knowledge of the order parameter, and in an extremely numerically cheap unsupervised manner. Now, we will show how it can be extended onto the quantum case and help reconstruct quantum phase diagrams of many-body systems from simple projective measurements. We will be using the transverse-field Ising and the Shastry-Sutherland models as examples.

The simplest example of a quantum phase transition is the paramagnet-to-ferromagnet transitions in the ferromagnetic Ising model in the transverse magnetic field given by the Hamiltonian
\begin{eqnarray}
H = J \sum_{ij} {\hat S}^z_{i} {\hat S}^z_{j} + h \sum_{i} {\hat S}^x_{i},
\end{eqnarray}
where $J$ and $h$ are the exchange interaction between nearest neighbour spins and the external magnetic field along $x$-axis, respectively, and we consider the case of one-dimensional chain with periodic boundaries. The critical value of magnetic field is known to be $h_c=0.5|J|$, and to reproduce this value is the first benchmark test for our method before we consider more sophisticated examples.

In the regime of weak magnetic field, the system's ground state obtained with the exact diagonalization approach \cite{Tom} is a superposition of two fully polarized states $\ket{\uparrow}^{\otimes N}$ and $\ket{\downarrow}^{\otimes N}$, which is nothing but the entangled GHZ state discussed above. In the $\sigma^z$ basis, the bit-string array generated by projective measurements is a random sequence of $000...0$ and $111...1$ blocks. In turn, at very high magnetic fields the qubits are pointing in the same direction along $x$ axis, and the state is just a trivial product state that can be obtained from $|0000...0\rangle$ by rotating all the qubits with the same Hadamard gate.

Fig.\ref{fig:Ising} shows the overall dissimilarity as a function of the magnetic field. One can see that in both $\sigma_z$- and random bases, the dissimilarity steadily decreases with increasing $h$, and the corresponding derivative ${\cal D}'(h)$ reveals the well-known transition point at $h=0.5$ (we take $J$=-1). The phase transition is also reflected in the 
partial dissimilarities ${\cal D}_k$ corresponding to individual renormalization steps. At low magnetic fields, the state is close to GHZ and there is clearly little inter-scale dissimilarity at small $k$: on the fine scale, coarse-graining of $|0000...0\rangle$ does not bring any dissimilarity, -- and the main contributions to $\cal D$ come from larger $k$, i.e. from the spatial scales covering several $N$-qubit blocks. Contrary to that, at larger fields finer scales start playing more important role. For each $k$, the phase transition at $h=0.5$ is visible in the derivative ${\cal D}'_k(h)$.

\begin{figure}[!ht]
	\includegraphics[width=1\columnwidth]{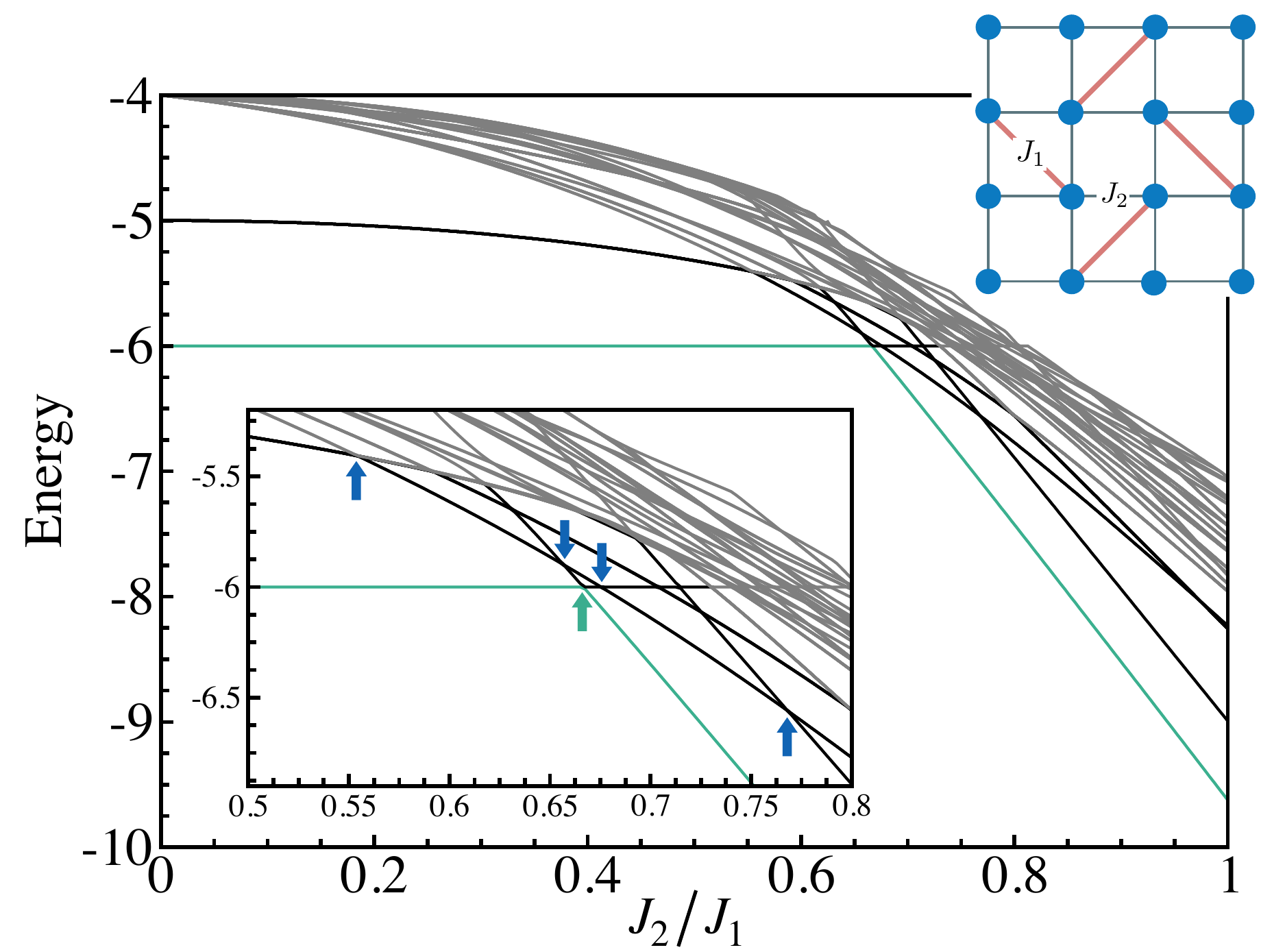}
	\caption{\label{fig:Shastry} Upper right inset: schematic representation of the Shastry-Sutherland model 16-spin supercell used in this work. Main plot and the inner inset: low-energy part of its spectrum as a function of the inter-dimer exchange interaction $J_2/J_1$. Arrows highlight transitions between quantum states. The green line represents the ground state.}
\end{figure}

\begin{figure}[!hb]
	\includegraphics[width=1\columnwidth]{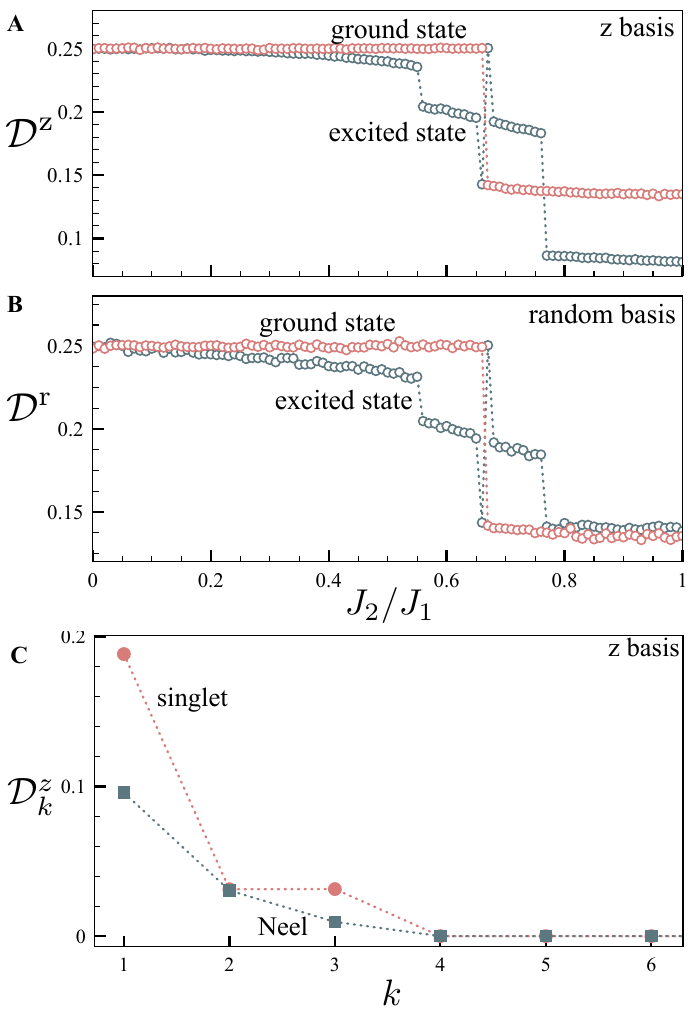}
	\caption{\label{fig:Shastry_dissimilarity}Dissimilarity of the ground and the first excited states of the Shastry-Sutherland model as a function of the inter-dimer coupling in the $\sigma^z$ ({\bf A}) and the random ({\bf B}) bases. ({\bf C}) Comparison of the partial dissimilarity profiles obtained for the singlet ($J_2$ = 0) and the N\'eel ($J_2$ = 1) states in the $\sigma^z$ basis.}
\end{figure}

A much less trivial test of the method is to check whether it can reveal transition points in highly-frustrated spin systems with richer phase diagrams. For that, we consider the Shastry-Sutherland model\cite{Shastry} with competing antiferromagnetic interactions on the orthogonal dimer lattice, which plays a crucial role in understanding physical properties of the SrCu(BO$_3$)$_2$ system \cite{Mila, Mila1, Mazurenko, Danis}. The corresponding Hamiltonian contains intra- and inter-dimer interactions, which are denoted $J_1$ and $J_2$ correspondingly (Fig.\ref{fig:Shastry}):
\begin{eqnarray}
H = \sum_{\rm dimer} J_1 \hat {\bf S}_i \hat {\bf S}_j + \sum_{\rm inter-dimer} J_2 \hat {\bf S}_i \hat {\bf S}_j. 
\end{eqnarray}
As it was previously shown, the system features a gapped singlet ground state at $J_2$ = 0, gapless long-range antiferromagnetic N\'eel state at $J_2 \gg J_1$, but also a plaquette phase in-between, in the range of $0.67 < J_2$/$J_1 < 0.76$. While, strictly speaking, the quantum phase transition is defined in the thermodynamics limit of infinite lattices, its precursor could be detected already in a small system \cite{Bagrov}. For example, in the case of Shastry-Sutherland model it has been suggested that by analyzing spin gap and spin-spin correlation functions one can extract the singlet-plaquette and plaquette-N\'eel transitions from exact diagonalization studies of small clusters \cite{Nakano}.
We are going to show that it can also be done with the inter-scale dissimilarity measure, which is agnostic about the nature of phase transition and much easier to implement on quantum simulators and quantum computers.

We have performed exact diagonalization study \cite{Tom} of a 16-spin Shastry-Sutherland supercell -- the smallest cluster on which the model can be defined. Its energy spectrum is presented in Fig.\ref{fig:Shastry}. One can see that up to $J_2 = 0.66J_1$ the ground state of the system is the singlet state separated from the first excited state with a non-zero spin gap, and its energy is independent on the inter-dimer coupling value $J_2$. At $J_2 = 0.66J_1$ a quantum phase transition takes place. The previous studies \cite{Nakano} have shown that increasing the supercell size does not change the position of the critical point. The inter-scale dissimilarity naturally captures this transition: for $J_2<0.66J_1$, $\cal D$ of the ground state computed from 8192 measurements is a constant, ${\cal D}=0.25$, and an abrupt transition occurs at the critical point in both the $\sigma^z$ and the random bases. The corresponding partial dissimilarities at $J_2=0$ and $J_2=J_1$ are shown in Fig.\ref{fig:Shastry_dissimilarity}.  

In the thermodynamic limit, the cases of $J_2 = 0$ and $J_2 = 1$ correspond to the magnetic phases with and without spin gap between the ground and the first excited state. In the finite-size system, it means that non-trivial signatures of phase transitions could be encoded not only in the ground state, but also in the excitation spectrum. At $J_2<0.55J_1$, the first excited state has three-fold kind degeneracy: it is of triplet type with total spin values $S^z$ = 0, $\pm 1$. Above the transition point, it is replaced with a two-fold degenerate state with zero total spin. This state reconfiguration causes the difference in magnetization profiles for the inter-dimer order parameter above and below the point of $J_2=0.55J_1$ when the external magnetic field is applied. According to the previous studies \cite{Miyahara}, the magnetization features a plateau at 1/8 of the full moment for $J_2$=0.65, but not for $J_2$=0.4.  

At the point of $J_2=0.76J_1$ (Fig.\ref{fig:Shastry}), the plaquette-N\'eel phase transition take place. Stability of this point upon varying the system size was previously confirmed by different methods \cite{Mila, Nakano, Lou}.

From Fig.\ref{fig:Shastry_dissimilarity}, one can see that all three transitions, -- at $J_2=0.55J_1$, $J_2=0.66J_1$, and $J_2=0.76J_1$, -- are accurately reflected in the inter-scale dissimilarity of bit-string array sampled in $\sigma_z$ and random bases from the first excited state of the Shastry-Sutherland model. We also show that the partial dissimilarities of the ground state calculated for $J_2$ = 0 and $J_2$ = 1 have specific distinguishable profiles.
We believe this to be a strong argument in favour of universality of the suggested approach to automatic construction of phase diagrams of many-body systems simulated on quantum devices.

So far we have been computing inter-scale dissimilarity of arrays composed out of 8192 measured bitstrings. However, it can be shown that in fact a much smaller number of measurements would suffice to complete the task of detecting phase transition points in many-body quantum systems. We found that, in the $\sigma^z$ basis, partial dissimilarities ${\cal D}_k$ of the Ising model ground states remain almost the same when we do 256 measurements instead of 8192. In the random basis, the minimal number of measurements that allows to reveal the ferromagnetic-paramagnetic transition is about 1024. In turn, the abrupt changes in the inter-scale dissimilarity of the Shastry-Sutherland model states could be revealed with mere 16 measurements. Thus, the method we propose allows one to accurately reconstruct phase diagrams of quantum spin Hamiltonians by using small-size supercells and a limited number of measurements.

\subsection{Multi-basis dissimilarity map}
\label{sec:IID}
So far, we have analyzed a number of distinct examples of quantum states and demonstrated that their inter-scale dissimilarities (both overall and partial) computed in different measurement bases can be regarded as easily measurable signatures. To make this discussion more concise, it is natural to consider all the states within a single unifying context.
 
\begin{figure}[!h]
	\includegraphics[width=1\columnwidth]{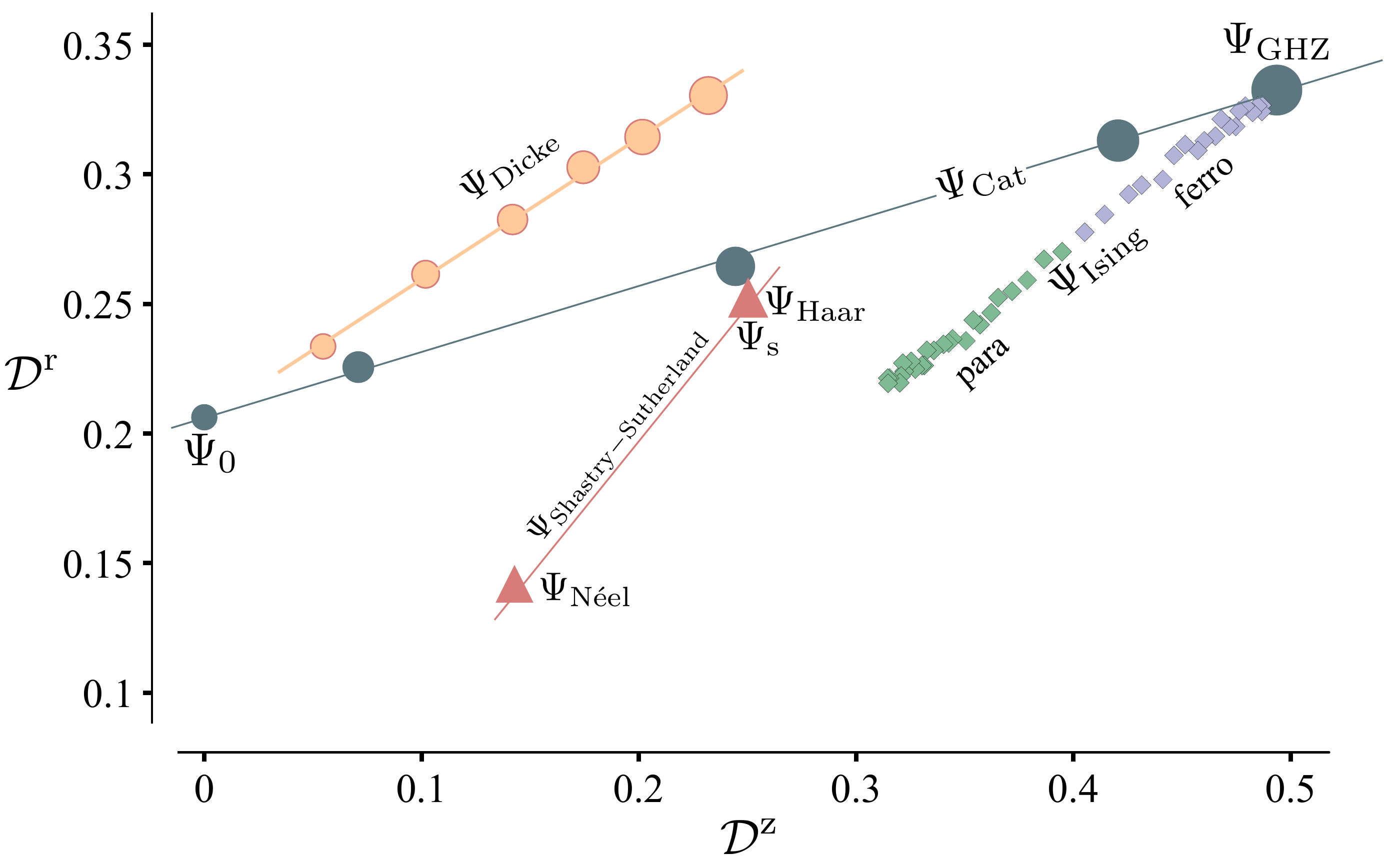}
	\caption{\label{fig:dmap} Dissimilarity map of the 16-qubit quantum states studied in this work. $\Psi_{\rm 0}$, $\Psi_{\rm s}$, $\Psi_{\rm Haar}$ denote the trivial $|0\rangle^{\otimes N}$, the singlet and the random quantum states, respectively.}
\end{figure}

To accomplish that, we shall introduce the concept of dissimilarity map. For the sake of nicer visualization, assume that we characterize each quantum state with only two numbers -- its overall dissimilarities ${\cal D}^z$ and ${\cal D}^r$ measured in the $\sigma_z$ and random bases correspondingly. Each state is then represented by a point in two-dimensional space. Fig.\ref{fig:dmap} shows several classes of states plotted on such a map. One can see that states belonging to different families nicely group in recognizable lines. The dissimilarity map can be then thought of as an approach to dimensional reduction that embeds higher-dimensional data in a plane (if more bases were used, it would be a three- or four-dimensional space instead). Some states still share the same location on the map, like the singlet state of the Shastry-Sutherland model $\Psi_{\rm s}$ and the chaotic state $\Psi_{\rm Haar}$ both having ${\cal D}^r$ = ${\cal D}^z$ = 0.25. This is not unexpected, since a many-body state cannot be uniquely represented with only two numbers. However, taking into account also their partial dissimilarity profiles (Figs.\ref{fig:Chaos} and \ref{fig:Shastry_dissimilarity}) we can distinguish the states. This way, $\cal D$ and ${\cal D}_k$ computed in several (two or more) different bases altogether form a hash of quantum state.

\section{Discussions}
In this paper, we have shown that bit-string arrays resulting from projective measurements of many-body quantum systems should be viewed as objects possessing internal hidden structure that contains important information about the measured quantum state. By computing inter-scale dissimilarities of the arrays, it is possible to define a specific characteristic of the state which serves as its ``hash'' that can be then used to certify the state and to estimate its closeness to the desired target state.

Two measures have been introduced: the overall dissimilarity $\cal D$ of the array in a chosen measurement basis, and the scale-dependent set of partial dissimilarities ${\cal D}_k$, which are building blocks of the quantum state signature. Since the bit-string array in a fixed basis is defined only by the probability distribution over the Hilbert space basis $|\psi(S_i)|^2$, it does not distinguish between two wave functions with the same set of amplitudes but different structure of phase. Thus it is important to compute $\cal D$ and ${\cal D}_k$ in two or more different bases. Since the procedure of performing projective measurements and computing dissimilarities is experimentally simple and numerically cheap, it is easy to repeat this procedure in several bases and construct a hash consisting of several numbers.

We would like to stress out that, in fact, the use of at least {\it two} measurement bases to characterize quantum system is not only practical, but also an important conceptual requirement directly related to Bohr's complementarity principle \cite{Bohr,Wheeler}. According to this principle, when observing a quantum system one gains information not about the quantum state {\it per se} but rather about the results of its interaction with a classical measuring device. Formally, the result of this interaction is described by the von Neumann theory of measurements \cite{vonneumann} as a projection of the system density matrix with only diagonal elements surviving in the basis dictated by the device. The use of at least two noncommutative projection operators corresponding to two complementary measurement devices is a necessary prerequisite of quantumness, as follows from a general ``separation-of-conditions principle'' \cite{separation}. The latter dictates a description of quantum quantities by, at least, two-index matrices rather than ``classical'' strings.

It has to be admitted that uniqueness of this signature is not guaranteed, and one can not exclude the possibility that two distinct quantum states have similar sets of $\cal D$ and ${\cal D}_k$. However, if the number of involved measurement bases is large enough, such a coincidence seems highly unlikely. 
Here, we have constructed merely two-dimensional dissimilarity maps for bit-string arrays obtained from measurements in the random and $\sigma^z$ bases, and this was already enough to characterize several important families of many-body quantum states. In the cases, when two different wave functions were indistinguishable on the map (like the singlet and the chaotic states), they could be distinguished by their ${\cal D}_k$ sets.
If one is concerned about issue of non-uniqueness, the method can be used as a cheap preprocessing scheme within a larger framework of certification. First the dissimilarity signature is computed, and if it strongly deviates from the target state signature, the prepared state can be discarded right away. And only if the two states appear close enough, more advanced analysis should be performed.

An important advantage of the proposed approach is its scalability. Due to simplicity of computing the inter-scale dissimilarities, this procedure can be conducted for a large number of qubits. By using a classical computer, one could potentially characterize states of quantum systems of several thousands qubits which goes far beyond the abilities of available intermediate-scale quantum devices. For example, if one uses 128 Gb RAM, the estimated sizes of quantum systems that can be characterized in this way lie in the range from 8192 to 1048576 qubits, if the number of bitstrings in the array is taken to be $2^{20}$ or $2^{13}$, correspondingly.

In this paper, we have analyzed two potential applications of the inter-scale dissimilarity signature, -- certification of quantum states and construction of phase diagrams. However, other research lines can be initiated, and we would like to briefly discuss them.

An important problem in quantum computing is to devise a quantum circuit that represents the desired target state. Usually, it is accomplished by optimization of the circuit architecture (topology, choice of gates) with overlap between the circuit and the target wave function being the objective function. For a large number of qubits, computing overlap at every iteration of optimization could be quite costly. Instead, one can aim at achieving the desired dissimilarity signature ${\cal D}_{target}$ and minimize the norm $||{\cal D}_{target} - {\cal D}_{circuit}||$ which, as discussed before, does not require significant resources to be computed even for a large system.

Another possible application of this concept could be in the domain of quantum optics experiments in which observer's eyes play the role of photons detector \cite{eye1,eye2} with a minimal detection threshold of single photon\cite{eye22}. Such a fascinating sensitivity of human eyes to the light has already become a basis for different scenarios of experiments \cite{eye3,eye4} aimed at detecting entanglement. Such experiments require accumulation of statistics over ``seen'' and ``not seen'' events. Since human eyes are much slower in counting light pulses than real photon detectors, collecting large amounts of data in such a setting is challenging, and a method that allows to harvest information from limited data could come handy. Representing two possible outcomes of a single measurement, ``seen'' or ``not seen'', as binary digits, one can construct an array that can be analyzed from the inter-scale dissimilarity point of view. As has been exemplified with Dicke and Schr\"odinger cat states, the latter can be used to estimate entanglement entropy of the state.

Finally, it should be highlighted that by constructing the low-dimensional dissimilarity map for a number of quantum states (as in Sec. II C) one, in fact, performs automatic dimensional reduction and visualization of a high-dimensional dataset -- a common task in machine learning which is often solved in unsupervised manner 
by employing such methods as self-organized Kohonen map, t-distributed stochastic neighbor embedding ($t$-SNE) \cite{sne,tsne}, or uniform manifold approximation and projection algorithms \cite{UMAP} (see Ref.\onlinecite{bagrov-UMAP} for a primer of how the latter can be used in the context of many-body quantum physics). These algorithms usually require some notion of distance between the original higher-dimensional data points and try to approximately preserve the relative distances when projecting points onto a lower-dimensional space (usually, two- or three-dimensional). By computing and visualizing dissimilarity signatures using two or three complementary measurement bases, one effectively solves the same problem for a dataset consisting of many-body quantum states. While it is possible to use the conventional dimensional reduction methods to classify and visualize quantum states by defining fidelity-based distance between them\cite{qt-SNE}, this would require storing and manipulating many-body states on a classical computer. Thus, using dissimilarity maps could be an easy to implement alternative that does not require much resources. Although it is not directly related to the distance between quantum states in the Hilbert space, it nevertheless consistently and neatly clusters quantum states belonging to different families without even relying on any optimization scheme.

\section{Methods}

\subsection{Calculating inter-scale dissimilarity of bit-string arrays}
To assign a characteristic hash function to a quantum state we perform three steps (Fig.\ref{fig:protocol}): (i) initialization of the quantum state on a real quantum device or simulator, (ii) a number of projective measurements in at least two different bases, and (iii) computing the inter-scale dissimilarities of the resulting bit-string arrays.

The {\it initialization} of a quantum state may be done by different means. For instance, one can use variational approaches \cite{Troyer,VQE, Sotnikov} and adiabatic algorithms \cite{adiab1,adiab2,adiab3} to approximate the target state on a quantum device. When dealing with a some small-scale quantum system, like the 16-qubit states studied in this paper, it is possible to initialize a state by taking the wave function coefficients obtained with exact diagonalization and employing the Least Significant Bit procedure \cite{LSB} that features one-by-one disentanglement of qubits. Some particular quantum states can be directly generated with known quantum circuits, which is the case for the quantum chaos and the Schr\"odinger cat states. In this work, all the manipulations with quantum states were performed with the Qiskit package \cite{IBM}.

Once a quantum state is initialized on a device, we measure it in two or more bases. Here, we refrained to projective measurements in the $\sigma^z$ basis and the random basis, though using more bases can be beneficial for constructing unique hashes of many-body states. 
In other words, we sample $N_{shots}$ basis vectors represented by bitstrings $\{ x_{i} \}$ from the probability distribution $p(x_{i})=|\psi(x_i)|^2$, where $N_{shots}$ is a reasonably small number of measurements (16 to 8192 in the studied cases), and by doing this in two bases we should have access not only to the amplitudes, but also to the phases of the wave function. The measurement outputs in each basis are then arranged into one-dimensional sequence of bitstrings which can be regarded as a binary array of length $L= N\times N_{shots}$.
Random basis measurements are performed in the following way. Prior to every shot $i$ of measurement, rotational gate $U^{(i)}_0$ parametrized by randomly generated angles $\theta_i$, $\phi_i$ and $\lambda_i$ is applied to each qubit (Fig.\ref{fig:protocol} A). For the next shot, new values $\theta_{i+1}$, $\phi_{i+1}$ and $\lambda_{i+1}$ are sampled, and a new rotational gate $U^{(i+1)}_0$ is applied. The angles are generated in such a way that, once the procedure is repeated many times, the single-shot gates uniformly cover a segment of the Bloch sphere: $\theta \in [0,\frac{\pi}{2}]$, $\phi \in [0,\frac{\pi}{2}]$ and $\lambda \in [0, \frac{\pi}{2}]$.
The reason why we choose one of the bases to be random in the aforedescribed sense is that it is expected to be the most unbiased one if we apply this protocol to diverse quantum states with completely different structures.

Having constructed the bit-string arrays, we analyze their structure using the concept of {\it inter-scale dissimilarity}. Recently\cite{complexity}, some of us have suggested a notion of structural complexity of classical patterns based on the idea of quantifying differences between distinct spatial scales of a pattern obtained with a multi-step renormalization (coarse-graining) protocol. Here, we formally apply this procedure to the bit-string arrays viewing them as one-dimensional patterns.

Let us denote such an array as vector ${\bf b}^0$ of length $L$. At every step of coarse-graining $k$, a vector of the same length is constructed as
\begin{gather}
b_{i}^{k} = \frac{1}{\Lambda^k} \sum_{l=1}^{\Lambda^k} b_{\Lambda^k[(i-1)/\Lambda^k]+l}^{k-1},
\label{eq:b_i^k}
\end{gather}
where square brackets denote taking integer part. This means that at each iteration the whole array is divided into blocks of $\Lambda^k$ size, and elements within a block are substituted with the same value resulting from averaging all elements of the block.  Initially those elements are either $0$ or $1$, and for $k>0$ they take real values (in fact, for the sake of nicer normalization in our calculations we assumed that ``$0$'' bits have values equal to $-1$). Index $l$ enumerates elements belonging to the same block. For simplicity, we usually assume that the bit-string length is an integer power of filter size $\Lambda$: $\log_{\Lambda} N \in {\mathbb N}$.

Dissimilarity between scales $k$ and $k+1$ is then defined as
\begin{gather}
\label{eq:Dissimilarity}
 {\cal D}_k =  |O_{k+1,k} - \frac12\left(O_{k, k}+O_{k+1, k+1}\right)|,
\end{gather}
where $O_{m,n}$ is the overlap between vectors at scales $m$ and $n$:
\begin{gather}
O_{m, n} = \frac{1}{L}\left({\bf b}^{m} \cdot {\bf b}^{n}\right).
\label{eq:overlap}
\end{gather}

There are two quantities of our principal interest: ${\cal D}_k$ that contains scale-resolved information on the pattern structure of the generated bit-string array and overall dissimilarity, ${\cal D} = \sum\limits_{k} {\cal D}_k$, where the sum goes over all the renormalization steps. $\cal D$ and $\left\{{\cal D}_k \right\}$ computed in several bases together comprise the hash function of quantum state that can be used for its certification.

\subsection{Dissimilarity of the random quantum state: analytical derivation}
Inter-scale dissimilarity of bit-string arrays resulting from projective measurements of random quantum states Eq. \eqref{eq:Dk_fit} can be estimated analytically. First, let us note that $O_{k,k}=O_{k,k-1}$ if the averaging-based coarse-graining scheme \eqref{eq:b_i^k} is adopted. Indeed, within $n$-th window of size $\Lambda^k$:
\begin{gather}
\frac{1}{\Lambda^k} \sum_{i=1}^{\Lambda^k} b_{\Lambda^k(n-1)+i}^{k} \cdot  b_{\Lambda^k(n-1)+i}^{k}=\\b_{\Lambda^k(n-1)+i}^{k} \cdot  b_{\Lambda^k(n-1)+i}^{k} \nonumber=\\b_{\Lambda^k(n-1)+i}^{k} \cdot \frac{1}{\Lambda^k} \sum_{i=1}^{\Lambda^k} b_{\Lambda^k(n-1)+i}^{k-1} \nonumber = \\ \nonumber \frac{1}{\Lambda^k} \sum_{i=1}^{\Lambda^k} b_{\Lambda^k(n-1)+i}^{k} \cdot  b_{\Lambda^k(n-1)+i}^{k-1},
\end{gather}
where $b_{(n-1)\cdot\Lambda^k+i}^{k}$ are equal to each other for all $i$ within the window, and thus this multiplier can be taken out of the sum over $i$.
Once summed up over all windows, l.h.s. of this identity gives $O_{k,k}$, and the r.h.s. -- $O_{k,k-1}$.

Thus, the expression for partial dissimilarity ${\cal D}_{k}$ can be rewritten as
\begin{equation}
    {\cal D}_{k}=\frac{1}{2}|O_{k+1, k+1}-O_{k, k}|.
\end{equation}
For a random state, $O_{k,k}$ can be evaluated in the assumption that binary elements in the bit-string array ${\bf b}_{i}^{0}$ are sampled from some random distribution $p_0(x)$ (with $x=0\, \mbox{or}\, 1$) and not correlated. In this case, the coarse-graining procedure can be viewed as follows. In step $k=1$, the renormalized probability distribution at every position in the array is defined over $x_1=0, \,0.5,\, 1$ with $p_1(0)=p^2_0(0)$, $p_1(0.5)=2p_0(0)p_0(1)$, $p_1(1)=p^2_0(1)$. Repeating this for several steps, one can notice that probability distribution $p_k(x_k)$ is defined over random variables which are obtained by averaging of the original uncorrelated random variables $x$, and according to the central limit theorem $p_k\to \mathcal{N}(\mu, \sigma^2/\Lambda^{k})$ as $k\to \infty$. Here $\mathcal{N}(\mu, \sigma^2/\Lambda^{k})(x)$ is a normal distribution with $\mu$ and $\sigma^2$ being the mean and variance of the original distribution $p_0(x)$ correspondingly, and normalization factor $1/\Lambda^{k}$ is due to the used scheme of averaging. 

Noticing that, on average, product of a site value on itself is
\begin{equation}
\langle (b_i^k)^2 \rangle_i = \frac{1}{L} \sum_{i=1}^{L}(b_{i}^{k})^2 \simeq \int x^2 p_k(x)dx,
\end{equation}
where the integral symbolically denotes discrete finite sum at finite $k$, we can approximately rewrite $O_{k,k}$ as:
\begin{equation}
O_{k,k}=\frac{1}{L}\sum_{i=1}^{L}(b_{i}^{k})^2 \simeq \int x^2 p_k(x)dx,    
\end{equation}
which leads us to
\begin{equation}
O_{k,k}\simeq\int x^2 \mathcal{N}(\mu, \sigma/\Lambda^{k})(x)dx=\mu^2+\frac{\sigma^2}{\Lambda^{k}}.
\end{equation}
 In this way, we obtain for $k>0$:
\begin{equation}
    {\cal D}_{k}=\frac{1}{2}[O_{k, k}-O_{k+1, k+1}]=\frac{\sigma^2}{2\Lambda^k}(1-\Lambda^{-1})
\end{equation}
Although the central limit theorem formally holds for $k\to\infty$, it turns out that this estimate reproduces the numerically computed partial dissimilarities already starting with $k=1$.

For $k=0$ it should be computed separately. Given $O_{0,0}\simeq \langle x^2\rangle$, we obtain:
\begin{equation}
{\cal D}_{0}\simeq\frac{1}{2}( \langle x^2\rangle-\mu^2-\frac{\sigma^2}{\Lambda})=\frac{\sigma^2}{2}(1-\Lambda^{-1})
\end{equation}

\section{Acknowledgements}
We thank Alexander Tsirlin and Tom Westerhout for useful discussions. The work of V.V.M., O.M.S. and I.A.I. was supported by the joint Russia-Swiss preparation grants. A.A.B. acknowledges financial  support  from  Knut  and  Alice  Wallenberg  Foundation through Grant No.2018.0060.  M.I.K. acknowledges a support from European Research Council via Synergy grant 854843 (FASTCORR). Exact diagonalization and quantum similator calculations were performed on the Uran supercomputer at the IMM UB RAS. We acknowledge the use of IBM Quantum services for this work. The views expressed are those of the authors, and do not reflect the official policy or position of IBM or the IBM Quantum team.

\end{document}